\documentstyle[11pt,cs12,html]{article}

\begin{document}
\title{Chromospheric~Dynamics~of~Betelgeuse~from~STIS~Spectra$^{1}$}
\author{Alex Lobel}
\affil{Harvard-Smithsonian Center for Astrophysics}

\footnotetext[1]{Based in part on observations with the NASA/ESA {Hubble Space Telescope} obtained at 
the Space Telescope Science Institute, which is operated by AURA, Inc., under NASA contract NAS 5-26555.}

\begin{abstract}
We present a high-resolution spectral analysis of Betelgeuse (M2 Iab). 
Between 1998 January and 1999 March four spatially resolved
\htmladdnormallink{raster scans}{http://cfa-www.harvard.edu/cfa/ep/pressrel/alobel0100.html}
have been obtained with the {\it STIS} on the {\it Hubble Space Telescope}.
The near-UV echelle spectra reveal double-peaked permitted emission lines of neutral and singly ionized 
metals, with self-absorbed line cores. We observe reversals in the intensity of both emission line 
components when scanning across the UV disk, for four unsaturated lines of Si~{\sc i}, 
Fe~{\sc ii}, Al~{\sc ii}], and Fe~{\sc ii}. We model the Si~{\sc i} $\lambda$2516 resonance 
line with detailed non-LTE radiative transport calculations in spherical geometry, and constrain 
the mean velocity structure in the projected aperture area, for each scan position on the chromospheric 
disk. We infer the spatial velocity structure of Betelgeuse's extended chromosphere, which reveals 
localized upflows in the western front hemisphere in 1998 September, that expand further toward the 
eastern hemisphere in 1999 March. The spatial scans exhibit simultaneous up- and downflows across the 
lower chromosphere with mean velocities of $\sim$2~$\rm km\,s^{-1}$. We infer non-radial (or non-coherent) 
mass movements during certain phases of the stellar variability cycle from these subsonic flows.
We present a discussion of constructing semi-empiric models for the chromosphere of this cool 
supergiant, and of its temporal variability. 
\end{abstract}

\section{Introduction: STIS and FOC Observations}

We have observed Betelgeuse's chromospheric disk four times with high 
spectral (E230M; R$\sim$30,000) and spatial resolution between 1998 January and 1999 March 
with the $HST$-STIS. This period of 15 months spans the photometric 
variability period of 400$-$420~d. observed for this cool supergiant.
The wavelength range of these echelle spectra spans from 2275~\AA\, 
to 3120~\AA. The spatial raster scans are obtained in steps 
of 25~mas across the UV-disk, providing S/N-values ranging from 40 at 
intensity peakup (Target Position\, `TP' 0.0) to 20 near the disk edge.  

Figure~1 shows the STIS aperture (25 by 100~mas) scan positions with respect to 
near-UV continuum images, simultaneously obtained with the Faint Object Camera (FOC) through 
a passband of $\sim$300~\AA\, centered at 2550~\AA. Betelgeuse's UV continuum chromosphere 
extends over ca. 120~mas, which is at least twice larger than its optical diameter, inferred from interferometric 
observations. These images of the chromosphere consist of four co-aligned dithered observations
obtained for pointing offsets of about half a pixel size (1 pixel = 14.35 mas). 
The image intensities are normalized to the maximum pixel to improve and to compare
the contrast of brighter (somewhat irregular) intensity patterns, mainly observed near the disk center.
The variable fluxes suggest small changes in the thermodynamic conditions of the 
lower chromosphere where the near-UV continuum forms. They can result from
small opacity changes caused by dynamic (and thermal) perturbations, originating from the 
deeper pulsating photosphere of this cool supergiant (for a discussion see Lobel \& Dupree 2000). 
We find that the detailed distributions of these brightness patterns remain 
currently unresolved with the spatial resolution of the FOC.   

\begin{figure}
\plotfiddle{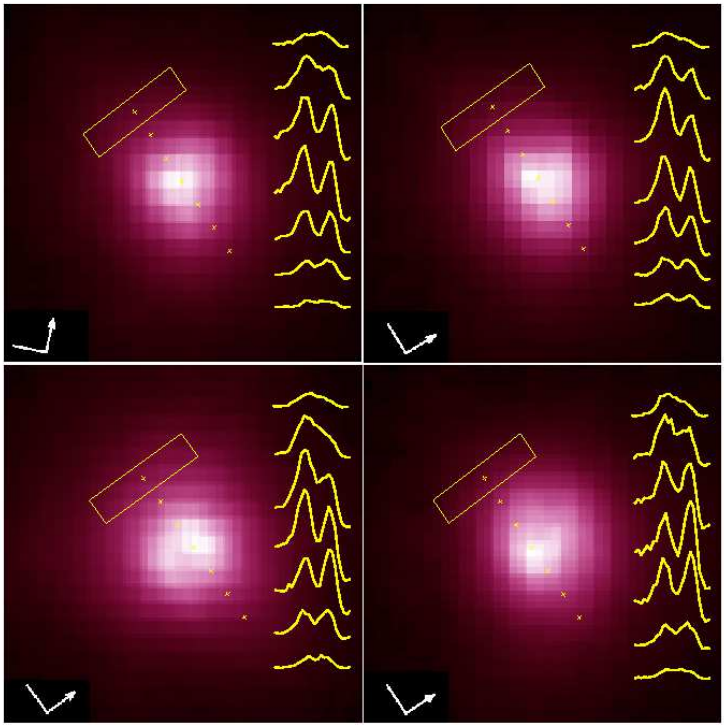}{5.5cm}{0}{50}{50}{-150}{-140}
\vspace*{4cm}
\caption{Near-UV images of Betelgeuse's chromosphere and simultaneous STIS raster 
scans, obtained in 1998 Jan. ({\it top left}), 1998 April ({\it top right}), 1998 Sept.
({\it bottom left}), and 1999 March ({\it bottom right}). Arrows point to north, east is left. }
\end{figure}
                 
Figure~1 shows the emergent intensity distribution of the $\lambda$2869 Fe~{\sc ii} line.
The profile of this chromospheric emission line is double-peaked due to a 
central scattering core. The asymmetries observed in these self-absorbed profiles 
of unsaturated (and unblended) lines serve as accurate indicators of the
velocity structure in their mean chromospheric line formation region.
In 1998 September ({\it lower left panel}) we observe a significant inversion in the intensity ratio 
of both emission line components when scanning from TP~0.0 to $+$0.025 
for Si~{\sc i} $\lambda$2516, Fe~{\sc ii} $\lambda$2869, 
Fe~{\sc ii} $\lambda$2402 and Al~{\sc ii}] $\lambda$2669, which we discuss in Section 4. 

\section{Atmospheric Model and Spectral Synthesis}

Figure 2 shows a comparison of low-resolution spectro-photometry ($\Delta$$\lambda$$\sim$25~\AA),
observed in the optical ({\it red and black lines}) with a detailed synthesis of the optical 
spectral energy distribution. The synthetic spectrum ({\it bold green line}) is computed with 
a photospheric model (Kurucz 1996) of $T_{\rm eff}$=3500~K, log($g$)=$-$0.5, 
$V_{\rm micro}$=2~$\rm km\,s^{-1}$, and has been broadened to a very low resolution. 
The local optical continuum level is considerably below the stellar continuum level 
({\it thin green line}), mainly due to enhanced opacity of TiO in this M2 star.  
The photospheric parameters have been determined by Lobel \& Dupree (2000a) based 
on unblended metal absorption lines in the near-IR, where TiO opacity strongly diminishes. 
We do not detect the He~{\sc i} 10830 \AA\, line in $\alpha$~Ori.

\begin{figure}
\plotfiddle{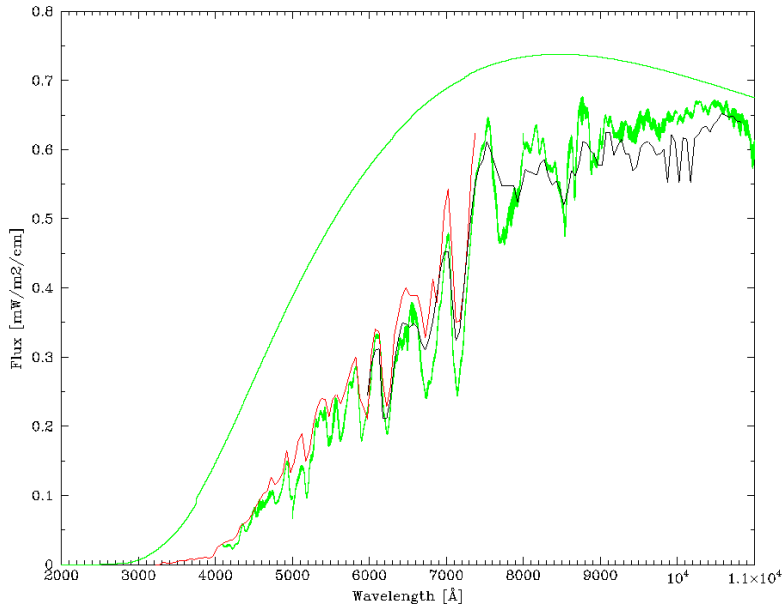}{5.5cm}{-90}{40}{40}{-160}{170}
\vspace*{2cm}
\caption{Observed ({\it black and red line}) and computed ({\it green lines}) 
optical spectral energy distribution of $\alpha$ Ori dominated by TiO opacity.
The stellar continuum is shown with the thin green line.}
\end{figure}

In Fig. 3 we compare a portion of the observed optical spectrum ({\it black line}) 
with the high-resolution synthetic spectrum ({\it red line}) computed in LTE. The model
correctly reproduces the relative intensities of prominent TiO band-heads, and requires
a (Gaussian) macrobroadening velocity of 12~$\rm km\,s^{-1}$ to fit the observed shape
of the spectral features. The green lines show the unbroadened spectrum.
We obtain the same broadening value from unblended lines 
in the near-IR, indicating supersonic large-scale mass movements in the deeper 
and higher layers of the photosphere. The best fits ({\it red lines}) to 
these photospheric lines, observed with NOT-Sofin ({\it black lines}), are shown in 
the upper panels of Fig. 4. The lower panel shows the best fit ({\it red line}) to the 
TiO bandhead at 6782~\AA, observed with WHT-UES between 1993-98. The band becomes too 
intense for the model with $T_{\rm eff}$=3600~K ({\it short-dashed line}) because 
the higher temperature enhances its contrast with respect to the overall TiO background, which becomes weaker.
The model with $T_{\rm eff}$=3200~K ({\it dash-dotted line}) yields a V~{\sc i} line 
around 6785.4~\AA\, that is too weak (after broadening) compared to the observed spectra. 

\begin{figure}
\plotfiddle{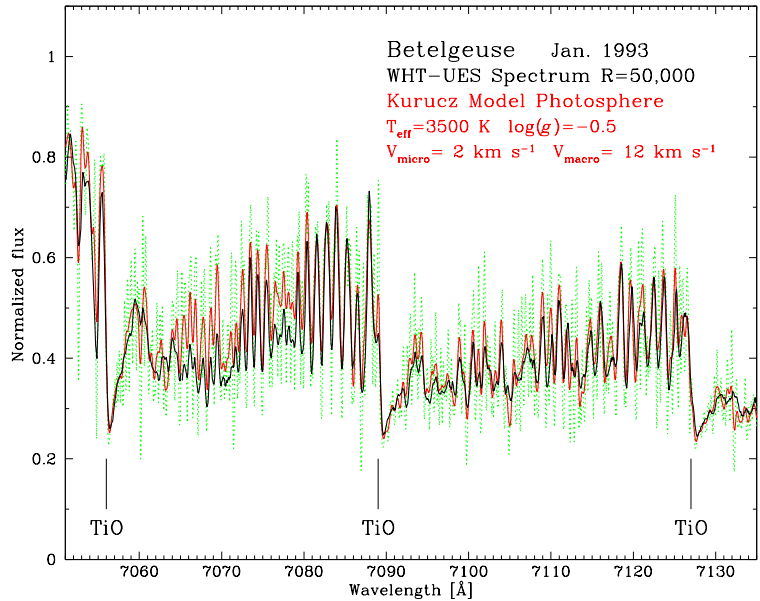}{5.5cm}{-90}{40}{40}{-160}{180}
\vspace*{2cm}
\caption{Observed ({\it black line}) and computed ({\it red line}) high-resolution optical 
spectrum of $\alpha$ Ori showing three TiO bandheads.}
\end{figure}   

\section{Detailed NLTE Line Profile Modeling}

The upper panel of Fig. 5 shows best fits to the H$\alpha$ line core ({\it boldly drawn lines}). 
The profiles are computed ({\it thin drawn lines}) in spherical geometry with the S-MULTI 
code (Harper 1992). The statistical equilibrium and rate equations are solved 
for a multi-level atom in a dynamic (1D) model of the atmosphere. This model
incorporates the thermodynamic model for the warmer chromosphere, which we semi-empirically       
determine from these detailed NLTE fits to the H$\alpha$ line depth and shape.
H$\alpha$ forms entirely in the chromosphere, which extends 
to $\sim$7~$R_{\star}$ above the photosphere, for $R_{\star}$=700~$R_{\odot}$.
Our best fits yield temperatures not in excess of 5500~K, 
and $N_{\rm e}$=1$-$7\,$\times$\,$10^{7}$~$\rm cm^{-3}$.
Note how the H$\alpha$ line core is distorted and appears 
asymmetric because of a hyperfine splitting band of photospheric Co~{\sc i} lines 
in its red wing.
The chromospheric temperature and electron density structure are shown 
in the middle panel of Fig. 5, determined for 1993 Feb. and 1996 May. 
Without a chromosphere the H$\alpha$ line core would remain invisible in the TiO 
background ({\it green dotted line}). The detailed fits to the shape and
equivalent width of H$\alpha$ also require supersonic microturbulence values,
ranging up to 19~$\rm km\,s^{-1}$, in the mean chromospheric formation region ({\it bottom panel of Fig. 5}). 
These values are in stark contrast with the small microturbulent 
velocity of only 2$\pm$1~$\rm km\,s^{-1}$ determined from the photospheric lines.
We constrain the chromospheric temperature minimum of $\sim$2700~K from detailed fits 
to the depth of the sharp TiO bandhead at 8860~\AA. We find clear observational 
evidence for intensity changes in the line depth of H$\alpha$, which correlate with   
constrast changes of the bandhead at 6782~\AA, observed between 1993-98 (Fig. 4). 
This indicates important long-term correlations between the thermal conditions of the extended 
chromosphere and of the upper photospheric layers near the temperature minimum, where the TiO bands form.
On the other hand, we could not detect significant changes in the depth 
and shape of the deep line cores of Fe~{\sc i} near 8868~\AA\, and 8870~\AA,
observed with UES on 1993 January 13 and February 12.   

\begin{figure}
\plotfiddle{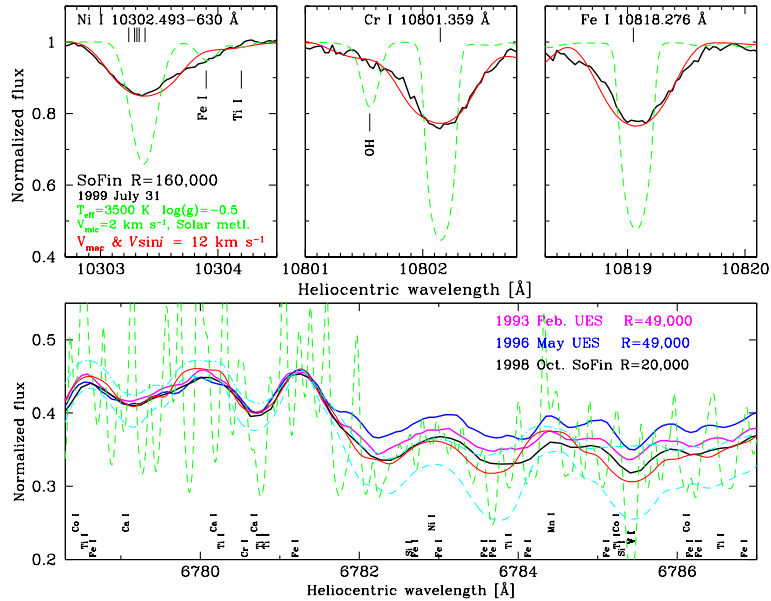}{5.5cm}{-90}{40}{40}{-160}{180}
\vspace*{2cm}
\caption{{\it Upper panels:} Best fits ({\it red lines}) to high resolution 
observations of photospheric metal lines ({\it black lines}) in $\alpha$~Ori. {\it Lower panel:}
Best fit ({\it red line}) to an optical TiO bandhead which reveals long-term intensity changes
that correlate with the intensity changes of H$\alpha$.}
\end{figure}

Figure 6 shows the detailed NLTE modeling of the Mg~{\sc ii} resonance doublet.    
We compute that the blue emission component of the $k$ line is blended with a strongly self-absorbed 
resonance line of Mn~{\sc i} $\lambda$2794.8. This narrow chromospheric scattering 
core (together with a blend of Fe~{\sc i}) strongly contributes to the
asymmetry of the emission components observed in the $h$ and $k$ lines of Betelgeuse
with STIS ({\it black line}) and GHRS ({\it blue line}). Another self-reversed emission line of 
Mn~{\sc i} blends with the blue emission component of the $h$ line. 
We compute that the Mg~{\sc ii} lines are strongly opacity sensitive which results from the large chromospheric 
column density. The strong variability observed in the red wing of their broad and saturated 
self-absorption cores indicates long-term changes in the global chromospheric 
dynamics, which can be linked with the long-term changes of the chromospheric conditions 
we infer from H$\alpha$. 

\begin{figure}
\plotfiddle{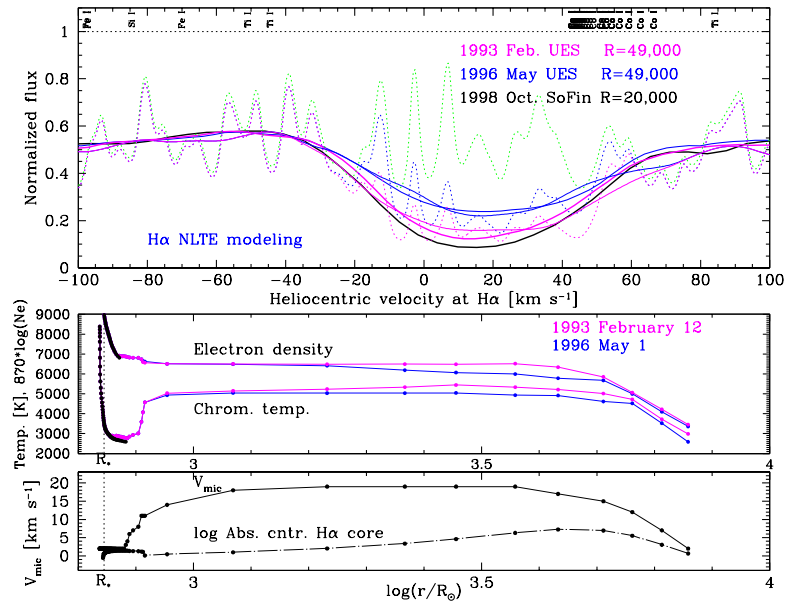}{5.5cm}{-90}{40}{40}{-160}{185}
\vspace*{1.5cm}
\caption{{\it Upper panel:} High resolution observations ({\it boldly drawn lines}) 
of H$\alpha$ in $\alpha$ Ori reveal long-term chromospheric variability. Best NLTE 
fits ({\it thin drawn lines}) constrain the kinetic temperature and electron density 
in the chromosphere ({\it middle panel}). {\it Bottom panel:} The detailed fits require 
a stong increase of the microturbulent velocity ($\rm V_{\rm mic}$) in the chromosphere, 
which extends over $\sim$7~$R_{\star}$.}
\end{figure}

The lower panel of Fig. 6 compares observations of the Ca~{\sc ii} K line core 
({\it black and blue lines}) with the detailed NLTE modeling of the line profile
with complete frequency redistribution (CRD, magenta line), and with 
partial frequency redistribution (PRD, red line) in a hydrostatic atmosphere. 
We find that the PRD calculation yields a better match to the observed 
line width of the central K emission core. However, we also find that the 
differences between CRD and PRD calculations remain limited for the 
much broader Mg~{\sc ii} resonance lines. These prominent emission lines 
reveal important effects caused by radiative transfer in spherical geometry 
compared to calculations in plane parallel geometry. In spherical geometry 
light rays that escape under high position angles travel a shorter distance through the 
chromosphere than through a flat slab of gas. This adds less thermal and turbulent 
broadening to the rays, which yields narrower line wings. In Betelgeuse it  
outweighs the Doppler diffusion into the line wings of Mg~{\sc ii}. For the 
weak Ca~{\sc ii} emission wings, which form closer to the chromospheric temperature 
increase, the turbulent broadening remains small, and PRD effects become 
appreciable for the line width.  

\begin{figure}
\plotfiddle{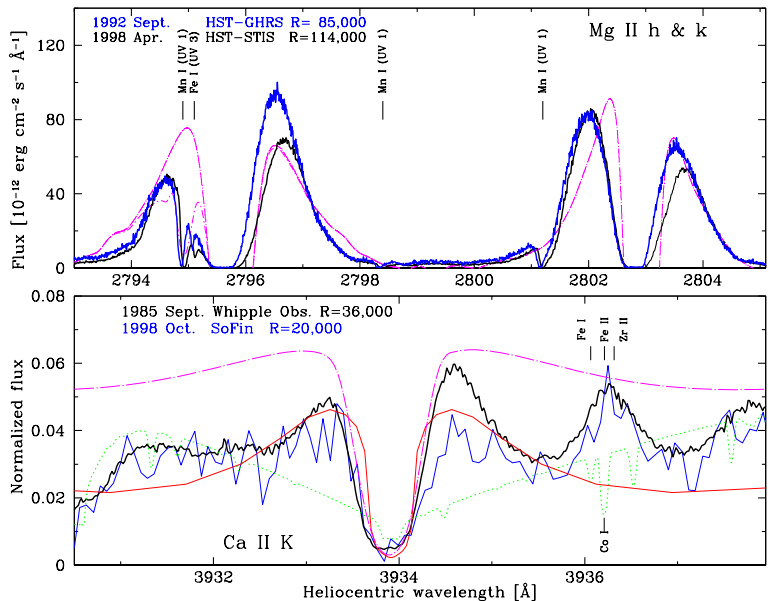}{5.5cm}{-90}{40}{40}{-160}{180}
\vspace*{1.6cm}
\caption{{\it Upper panel:} Variable Mg~{\sc ii} $h$ \& $k$ lines observed with high dispersion 
compared to detailed NLTE modeling ({\it magenta lines}). {\it Lower panel:} Ground-based 
observations of the Ca~{\sc ii} K line compared to CRD ({\it broken magenta line}) and PRD ({\it red line}) 
calculations in spherical geometry. The dotted line is computed in LTE without a chromosphere.}
\end{figure}

\begin{figure}
\plotfiddle{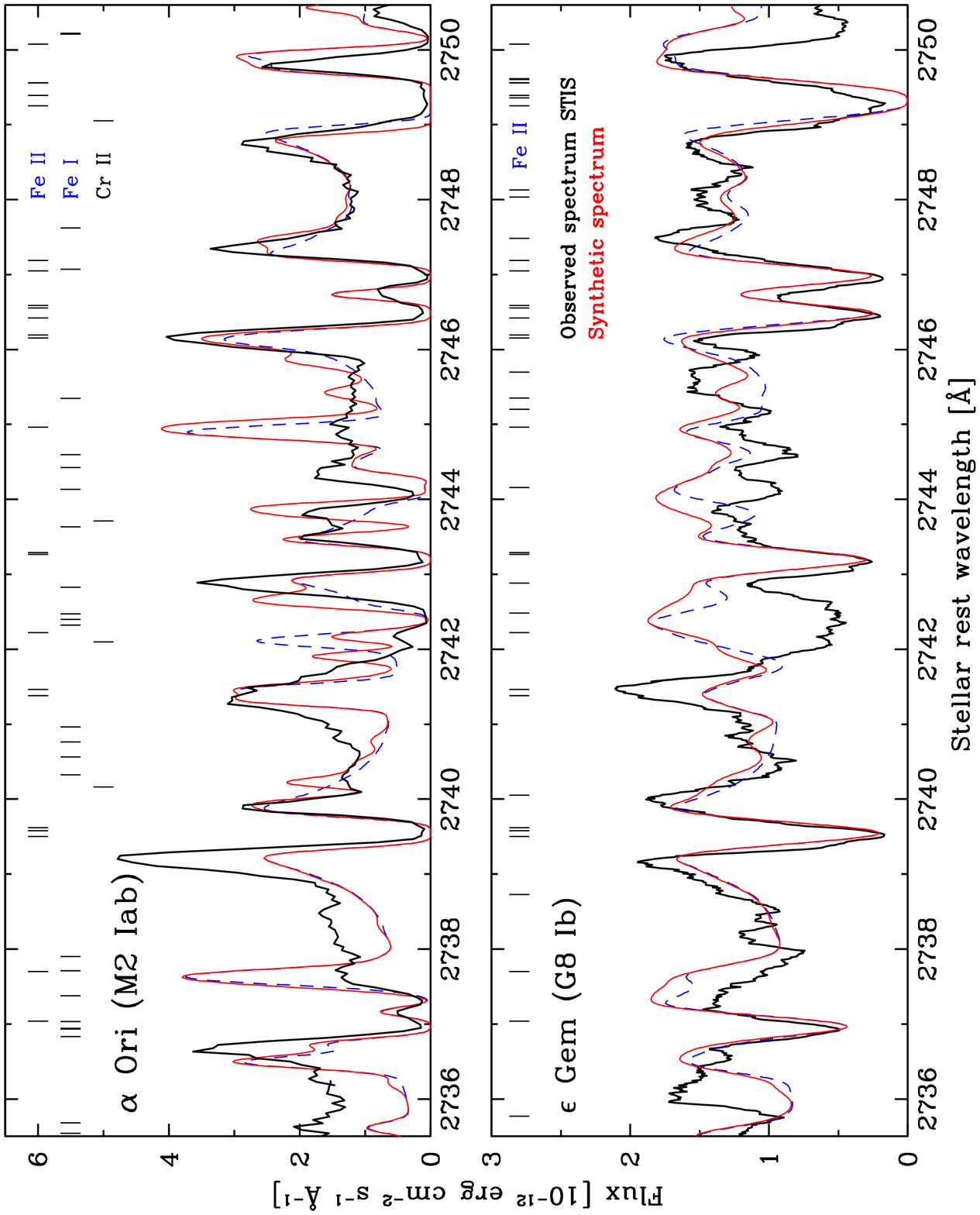}{5.5cm}{-90}{50}{40}{-200}{180}
\vspace*{2.cm}
\caption{Comparison of observed ({\it black lines}) and computed chromospheric 
spectra ({\it red lines use all atomic species}) for $\alpha$ Ori ({\it upper panel}) and $\epsilon$ Gem 
({\it lower panel}). The spectra are well reproduced with only iron lines ({\it blue dashed lines})
for a maximum chromospheric temperature of 5500~K in $\alpha$~Ori and 7050~K in $\epsilon$~Gem  }
\end{figure}

Figure 7 shows a portion of $\alpha$~Ori's near-UV spectrum observed with STIS 
at TP~0.0 in 1998 September ({\it black line}). The synthetic 
spectrum is computed with the mean thermodynamic 
model of Lobel \& Dupree (2000a). The majority of the 
strong absorption troughs in the spectrum are 
self-absorption cores of many blended Fe~{\sc i} and Fe~{\sc ii}  
lines ({\it dashed lines}). Our detailed (LTE) synthesis with all the atomic species 
also identifies a number weaker Cr~{\sc ii} lines 
({\it red line in upper panel}). The line list and the atomic line data are from Kurucz (1998),
which includes semi-empirical calculations of the oscillator strengths. 
The synthesis correctly reproduces the overall appearance of the spectrum, 
although it also yields unobserved strong iron emission lines
(i.e. Fe~{\sc i} $\lambda$2737.6 and Fe~{\sc ii} $\lambda$2744.9). The
log($gf$)-values of these lines are small and are computed to be too strong. 
We do not detect emission lines of ions higher than the first ionization stage in the STIS spectra
of $\alpha$~Ori.  
The lower panel of Fig. 7 shows the synthetic spectrum of the hotter 
supergiant $\epsilon$~Gem (G8 Ib). Its chromospheric model requires a 
maximum kinetic temperature of 7050 K (compared to 5500 K for $\alpha$ Ori), 
with a temperature minimum above 3100~K, to match the relative intensities of 
self-absorption cores with respect to the overall near-UV continuum.  
These cores are substantially deeper for $\alpha$ Ori due to the larger 
geometric extension (i.e. column density) of the chromosphere ($\sim$7~$R_{\star}$), 
whereas for $\epsilon$~Gem the chromospheric extension does not exceed $\sim$1~$R_{\star}$
(with $R_{\star}$$\simeq$140~$R_{\sun}$). Our modeling also requires 
a macrobroadening velocity of 9$\pm$1~$\rm km\,s^{-1}$ to match the chromospheric spectrum of $\alpha$~Ori, 
but which is larger for $\epsilon$~Gem, with $V_{\rm macro}$=21~$\rm km\,s^{-1}$.
These highly supersonic velocities reveal important large-scale mass movements in the chromospheres 
of cool supergiants, which cannot be attributed to fast stellar rotation (or large $v$sin$i$ values) for these 
evolved stars. Best fits to the near-UV spectra of earlier G-type supergiants as $\beta$~Dra (G2), and 
hybrid supergiants (as $\alpha$~Aqr and $\beta$~Aqr), also reveal supersonic macroturbulent 
velocities for their thinner chromospheres (Lobel \& Dupree 2000b).         

\section{Chromospheric Kinematics}

Figure 8 shows an overview of the emergent intensity distributions across 
the chromospheric disk (in scan offsets of 25 mas from intensity peakup TP 0.0) 
for 3 prominent emission lines of Si~{\sc i} $\lambda$2516 ({\it black line}), 
Fe~{\sc ii} $\lambda$2869 ({\it blue line}), and Al~{\sc ii} ] $\lambda$2669 ({\it green line}).  
For the four dates we observe that the stellar continuum emission reduces almost linearly 
from TP $+$0.025 to $+$0.075, about halving at each subsequent scan position. 
When scanning toward the limb this flux dimming for the bright emission lines,
is however less than for the stellar continuum level.
The red dashed lines show the (scaled) profiles of the Fe~{\sc ii} line obtained 
with the STIS high-resolution grating (R=114,000) on 1998 April 1. This 
spectrum is obtained with the larger 63 by 200 mas aperture, for 2 pointing offsets
of $\pm$63 and $\pm$126 mas (but shown here at $\pm$0.05 and $\pm$0.075 mas).
The comparison to the line shapes obtained with medium resolution (R$\simeq$30,000) 
reveals that the asymmetry and width of the profiles are not influenced by the 
spectral dispersion, and these medium resolution spectra are useful to infer 
the detailed chromospheric kinematics (i.e. its temporal and spatial variability).  

\begin{figure}
\plotfiddle{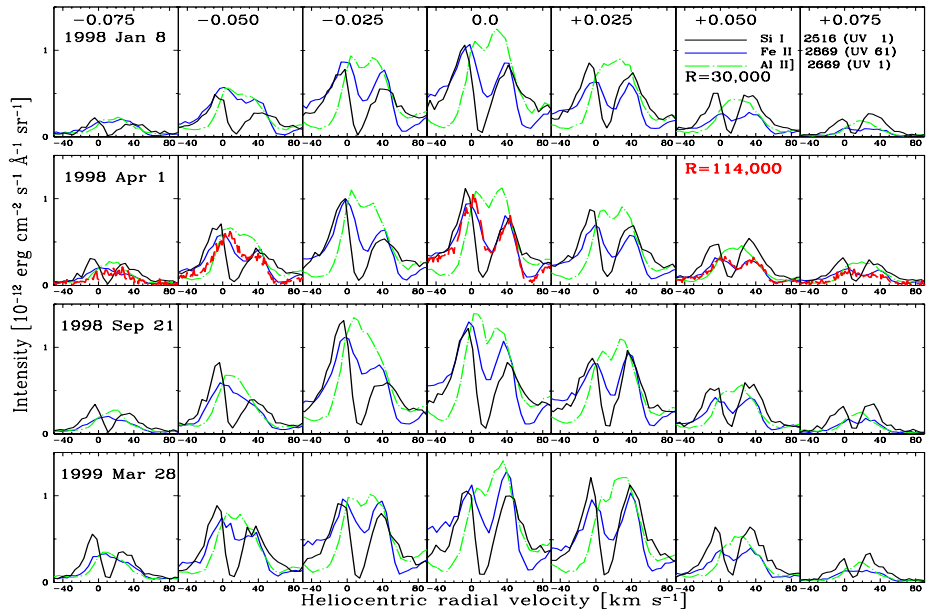}{5.5cm}{-90}{50}{40}{-200}{180}
\vspace*{2.cm}
\caption{Emergent intensity distributions of three unsaturated self-absorbed chromospheric 
emission lines of $\alpha$ Ori, obtained form {\it HST}-STIS raster scans for four observations dates
in 1998$-$99.}
\end{figure}

The raster scans of 1998 January and April display
profiles of prominently self-reversed emission lines with stronger 
short-wavelength emission maxima compared to their long-wavelength maxima. 
These optically thick emission lines reveal global downflows in the line formation region
across the chromosphere. The corresponding Doppler shifts 
by scattering material in front of the disk enhance the opacity in the 
red wing of the self-reversal, thereby suppressing the intensity 
of the red emission line component. In 1998 September we observe a remarkable 
reversal of the emission line asymmetry, scanning from TP 0.0 to TP $+$0.025.
This prominent reversal is observed in four unblended emission lines shown in Fig. 9.
These emission component intensity changes exceed the error bars 
of the line fluxes provided by the STIS calibration pipeline. 
In 1999 March we observe a further increase of the red emission 
component at TP 0.0, and the dimming of blue emission maxima for the negative TPs. 
It reveals systematic changes from a mean downflow into upflow over the line formation 
region of the central scattering cores. These simultaneous up- and 
downflows indicate global asymmetric (or nonradial) oscillations of 
Betelgeuse's extended chromosphere.                    

\begin{figure}
\plotfiddle{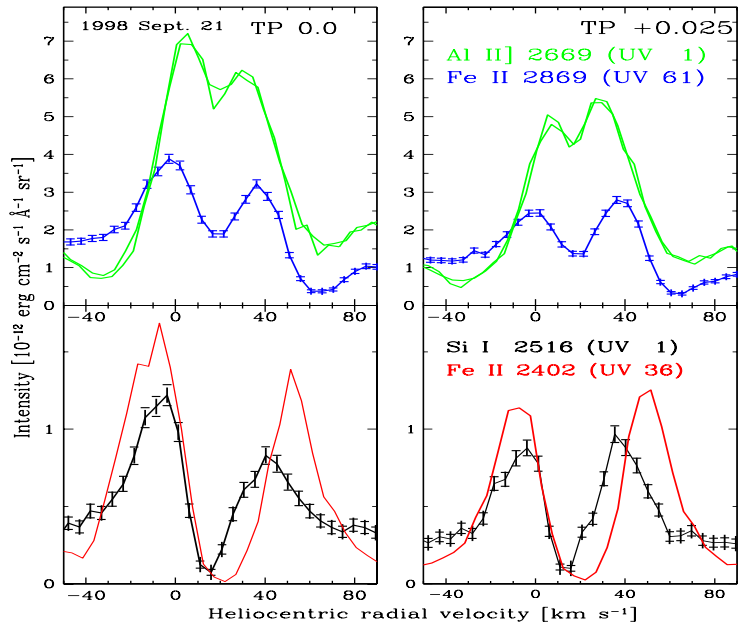}{5.5cm}{0}{50}{32}{-150}{-75}
\vspace*{2.1cm}
\caption{A prominent intensity reversal for the emission components of 
four self-absorbed lines is observed in 1998 September near the disk center (TP 0.0,{\it left panels};
TP $+$0.025,{\it right panels}) of $\alpha$~Ori. The Al~{\sc ii}~] line ({\it green lines}) 
is shown for two overlapping echelle orders, which reveals that the instrumental noise is small.}
\end{figure}

\begin{figure}
\plotfiddle{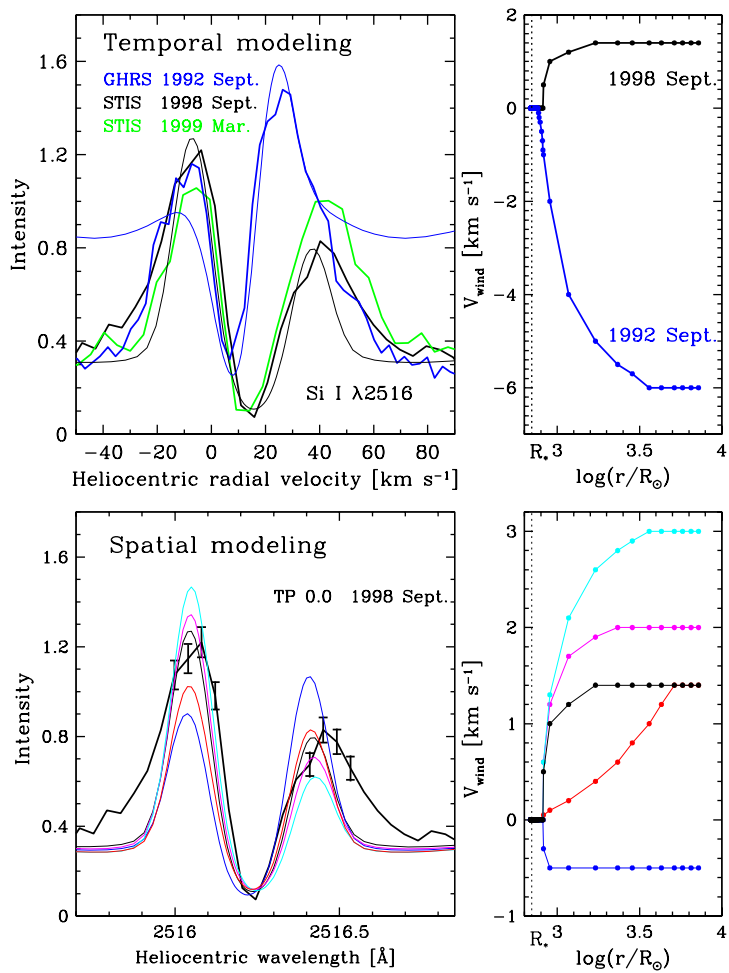}{5.5cm}{0}{50}{50}{-150}{-100}
\vspace*{3cm}
\caption{{\it Upper panels:} Temporal changes observed in the asymmetry of the 
Si~{\sc i} $\lambda$2516 resonance line ({\it boldly drawn lines}) reveal expanding (1992) and collapsing 
(1998) chromospheric velocity structures from best NLTE fits ({\it thin lines}).
{\it Lower panels:} Spatially resolved modeling ({\it thin lines}) of this line profile 
({\it bold black line shown for} TP 0.0) determines the velocity structure across the chromosphere 
for different scan positions, which reveals simultaneous in- and outflow in 1998 September.}
\end{figure}           

\begin{figure}
\plotfiddle{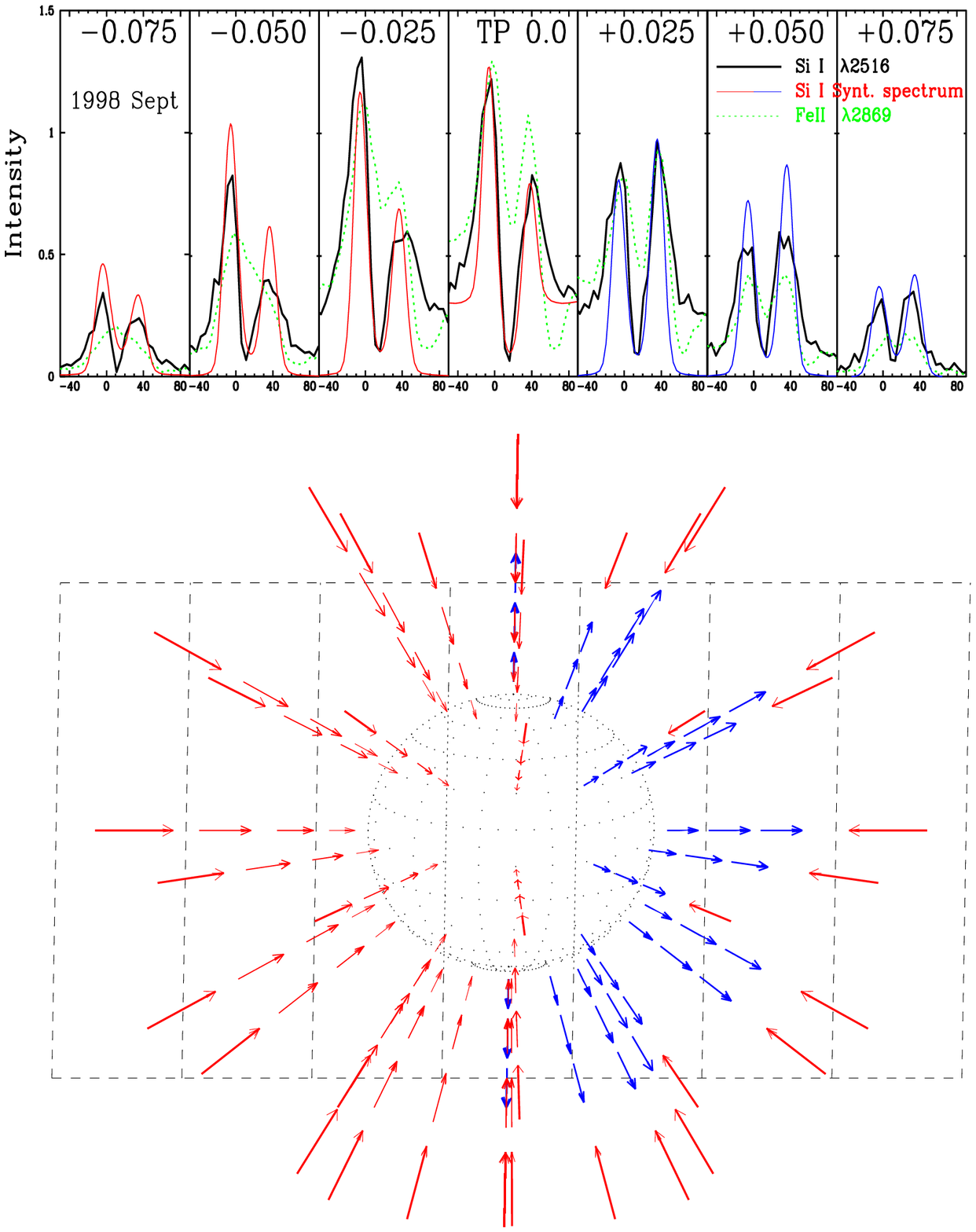}{5.5cm}{0}{60}{60}{-180}{-100}
\vspace*{3cm}
\caption{Schematic 3D representation of flow velocities in $\alpha$ Ori's chromosphere 
inferred from detailed fits ({\it thin lines}) to the asymmetric shape of Si~{\sc i} $\lambda$2516 
observed across the UV disk with STIS ({\it bold black lines}).
The arrows in the lower graph indicate the magnitude and the mean flow direction.
The emission line component intensity reversal from TP 0.0 to $+$0.025 reveals local upflow in the 
western front hemisphere (east is left) of the lower chromosphere in 1998 September.}
\end{figure}

\section{Detailed Dynamic Spectral Modeling}

The Si~{\sc i} $\lambda$2516 line also displays strong temporal changes besides
the spatial variations shown in Fig.~8. In the upper left panel of Fig.~10 
the (GHRS) disk-integrated self-absorption core of 1992 September reveals a blue-shift 
of $\sim$10~$\rm km\,s^{-1}$ ({\it bold blue line}). The red emission component strongly 
exceeds the blue one, while the latter becomes stronger in 1998 September ({\it bold black line}). 
We determine the chromospheric velocity structure by means of detailed radiative transfer fits.
We solve the statistical equilibrium for a Si~{\sc i} multi-level 
model atom, and compute emergent line profiles with our model atmosphere in spherical
geometry. The best profile fit is obtained ({\it thin blue line}) for a chromospheric velocity structure 
accelerating outwards to $-$6~$\rm km\,s^{-1}$ in 1992 ({\it upper panel right}).
Carpenter \& Robinson (1997) directly sampled this accelerating region. 
They detected in the scattering cores of 24 Fe~{\sc ii} lines a trend of 
increasing blueshift with increasing opacity or height. Our detailed modeling shows 
that the line shape of 1998 September is best fit ({\it thin black line}) with a collapsing 
velocity structure of $+$1.4~$\rm km\,s^{-1}$, which decelerates toward the photosphere. 
In 1999 September the scattering core shifts again blueward by $\sim$4~$\rm km\,s^{-1}$, 
enhancing the red emission component ({\it bold green line}). 

\begin{figure}
\plotfiddle{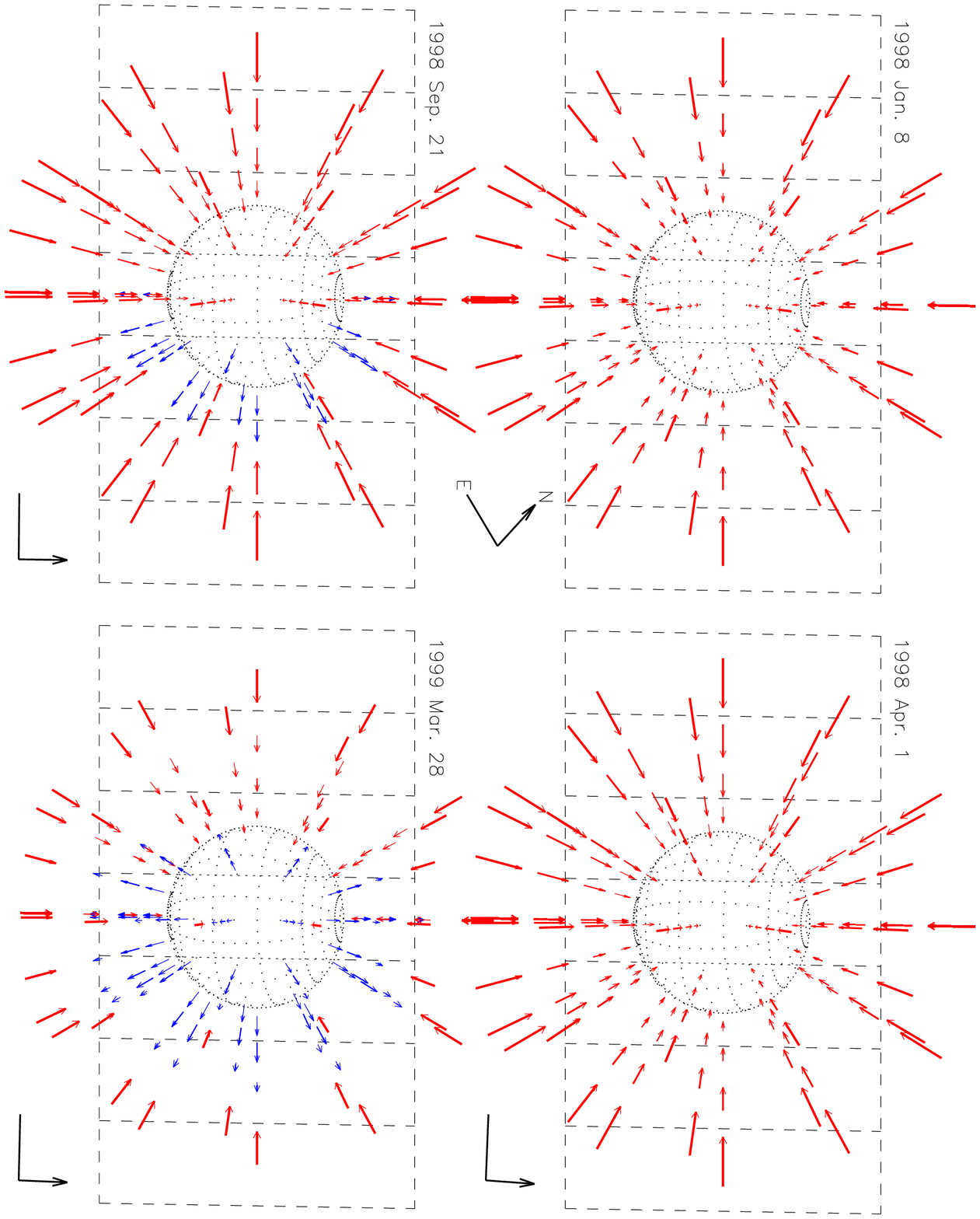}{5.5cm}{90}{60}{60}{250}{-30}
\vspace*{1.cm}
\caption{3D representation of fluid movements in $\alpha$ Ori's chromosphere inferred
from semiempirical modeling of Si~{\sc i} $\lambda$2516 in the four STIS raster scans. Other more 
optically thick lines define the dynamics at high levels. The photospheric radius 
is drawn by the dotted circles. The global downflow observed for the larger 
chromospheric envelope in 1998 January and April reverses into subsonic upflow for 1998 September 
in the lower chromospheric layers. The outflow enhances in 1999 March, when more 
symmetric profiles are observed ({\it bottom panel of Fig. 8}), extending farther toward 
the eastern front hemisphere.}
\end{figure}

We model the spatially resolved observations of Fig.~8 by means of spatially resolved 
radiative transfer calculations. Light rays which traverse the chromospheric model
are integrated along the width and height of the slit area, for each aperture position on the UV disk.       
The profile at TP 0.0 in the lower left panel of Fig.~10 ({\it bold black line}) is best fit 
({\it thin black line}) for the velocity structure shown in the lower right-hand panel ({\it black line}). 
However, the blue emission component becomes too weak for a slowly decelerating velocity
structure ({\it red lines}), whereas higher inflow velocities ({\it magenta and cyan lines}) 
suppress the long-wavelength emission line component too strongly. 
We determine that the reversal of component asymmetry in 1998 September between TP 0.0
and TP $+$0.025 of Fig.~9 corresponds to an outflow velocity in the lower chromosphere
of $-$0.5~$\rm km\,s^{-1}$ ({\it blue lines in lower panels of Fig. 10}). 
Hence, we find that the chromosphere assumes an inherently nonradial velocity structure 
in 1998 September. The lower panel of Fig.~11 
shows a schematic 3D-representation of the chromospheric kinematics, which collapses on average, 
but in which upflow occurs from the deeper layers at TP $+$0.025. 
We observe that this outflow extends farther
toward the eastern (left) front hemisphere in the raster scan of 1999 March (see Lobel \& Dupree 2001).
Notice how the line component asymmetry reduces near the limb due to the geometric 
projection of the chromospheric velocity field. This is also reproduced with our spatial radiative transfer
modeling. We measure a constant (Gaussian) macrobroadening of 9~$\rm km\,s^{-1}$ across the UV-disk, which 
is also observed in 1998 January when the scan axis was tilted by $-39{\deg}$ with respect to the east-west 
direction. This spatial invariance indicates that the marcobroadening is not determined by 
large rotation velocities ($v$sin$i$ values). This macrobroadening is rather linked with an isotropic 
and uniform chromospheric velocity field which cannot be interpreted as due to 
the spatial inhomogeneity of a large-scale granular field that consists of a few 
convection cells. 

\section{Chromospheric Nonradial Oscillation}

Our monitoring of the optical radial velocity curve of Betelgeuse at the Oak Ridge Observatory (CfA)
indicates semiregular pulsations of the photosphere over quasi-periods of 400$-$420 d. 
Optical long-term radial velocity curves provide amplitudes ranging to $\Delta$$v$$\sim$6~$\rm km\,s^{-1}$
(see Smith et al. 1989 for an overview). We observe a redshift of 4$-$5~$\rm km\,s^{-1}$
during the monitoring with STIS. An average collapse of the photosphere in the 
line of sight during this period is consistent with the mean downflow we observe for 
the larger chromospheric envelope. The localized upflow we detect in the deeper 
chromospheric layers can result from photospheric internal gravity waves that 
drive or interfere with the chromospheric kinematics. Other observational 
indications for local nonradial pulsation of $\alpha$~Ori have been provided by Hayes (1981).
Long-term monitoring of linear polarization changes do not conform with a single
{\it global} pulsation mode. However, {\it local} nonradial pulsations
provide a means of producing asymmetries that distort the star's spherical shape,
and which can account for the changes of net polarization. De Jager \& Eriksson (1992)
computed that a 440 day pulsation period can be interpreted as the period of internal gravity 
waves ($g$-modes) with a wavelength between $R_{\star}$/3 and $R_{\star}$/30. These 
low-order gravity waves ($l$$>$1) propagate nearly horizontally through the outer atmosphere.
They may arise above the convective layers by overshoot motions. When triggered 
in the very upper photospheric layers they can propagate randomly into the 
lower chromosphere and cause the local upflow movements we detect from the spatially
resolved STIS spectra.      

\section{Conclusions}

We find evidence for the presence of complex velocity fields
in Betelgeuse's extended chromosphere. Our finding is based on 
spatially resolved spectra obtained with the STIS with high spectral resolution
between 1998 January and 1999 March. We observe in the raster scan of 1998 September 
a prominent reversal in the intensity maxima of four self-absorbed
chromospheric emission lines near the disk center. Detailed (NLTE) radiative 
transport modeling of these optically thick emission lines 
reveals mean subsonic flows, streaming in opposite directions  
through the deeper chromosphere. The temporal behavior of the near-UV
emission lines indicates a chromospheric oscillation phase with 
global downflow between 1998 January and September. For the latter 
observation date, local upflow commenced in the lower chromosphere of 
the western front hemisphere, which extended farther toward the eastern 
hemisphere in 1999 March. We propose that the chromospheric kinematics 
is caused by nonradial or non-coherent oscillations during certain phases of the 
chromospheric variability cycle. Our detailed modeling reveals 
an increase of hydrodynamic velocities with the decrease of their 
length scales in Betelgeuse's chromosphere. The very large 
chromospheric oscillations are subsonic and nonradial. 
The large-scale macroturbulence is isotropic, with a velocity 
around the isothermal sound speed in the chromosphere (with $T_{\rm max}$=5500~K).
The small-scale microturbulent velocities are subsonic in the 
the photosphere and become highly supersonic in the chromosphere. 

\acknowledgments
I thank Drs. A. K. Dupree, R. Kurucz, E. Avrett, and H. Uitenbroek for useful discussions.
Dr. R. Gilliland is thanked for assistance with the {\it HST} proposal 7347 of the STIS 
observations. This research is supported in part by an STScI grant 
GO-5409.02-93A to the Smithsonian Astrophysical Observatory.

\index{*Betelgeuse|see {$\alpha$ Ori}}
\index{*HD 39801|see {$\alpha$ Ori}}
\index{*HR 2061|see {$\alpha$ Ori}}
\index{*58 Ori|see {$\alpha$ Ori}}
\index{*BD +7\deg1055|see {$\alpha$ Ori}}
\index{*SAO 113271|see {$\alpha$ Ori}}

\index{*HD159181| see {$\beta$ Dra}}
\index{*HD209750| see {$\alpha$ Aqr}}
\index{*HD204867| see {$\beta$ Aqr}}

\end{document}